\documentclass[12pt]{article}%
\usepackage{amsmath,latexsym}
\usepackage{graphicx}
\usepackage{amsmath}
\usepackage{amsfonts}
\usepackage{amssymb}
\begin{document}

\title{{Macroscopic traversable wormholes: minimum requirements}}
   \author{
Peter K.F. Kuhfittig*\\  \footnote{kuhfitti@msoe.edu}
 \small Department of Mathematics, Milwaukee School of
Engineering,\\
\small Milwaukee, Wisconsin 53202-3109, USA}

\date{}
 \maketitle

\begin{abstract}\noindent
While wormholes are just as good a 
prediction of Einstein's theory as 
black holes, they are subject to severe 
restrictions from quantum field theory. 
To allow for the possibility of interstellar 
travel, a macroscopic wormhole would need to 
maintain sufficiently low radial tidal forces.  
It is proposed in this paper that the 
assumption of zero tidal forces, i.e., the 
limiting case, is sufficient for overcoming
the restrictions from quantum field theory.  
The feasibility of this approach is subsequently 
discussed by (a) introducing the additional 
conditions needed to ensure that the radial 
tidal forces can indeed be sufficiently low
and (b)  by viewing traversable wormholes 
as emergent phenomena, thereby increasing 
the likelihood of their existence.
\end{abstract}

\section{Introduction}\label{S:introduction}

Wormholes have been a subject of interest ever 
since it was realized that the Schwarzschild  
solution and therefore black holes can be 
viewed as wormholes, albeit nontraversable. 
(The word ``wormhole" was coined by John 
Archibald Wheeler in the 1950's.)  More 
recently, the subject of entanglement has 
resulted in a renewed interest in a special 
type of wormhole, the Einstein-Rosen bridge, 
to explain how two particles can remain in 
contact even if they are widely separated 
\cite{ER35}.  A discussion of entangled black 
holes can be found in Ref. \cite{MS13}.  We 
will therefore assume that a basic wormhole 
structure can be hypothesized.

In this paper we are more concerned with 
macroscopic wormholes suitable for interstellar 
travel, first proposed by Morris and Thorne 
in 1988 \cite{MT88}, also discussed in Ref. 
\cite{MM21}.  While wormholes may be just 
as good a prediction of Einstein's theory as 
black holes, they are subject to severe 
restrictions from quantum field theory.  Here 
one of the biggest obstacles is the possible 
existence of large radial tidal forces, just 
as they are for black holes, leading to what 
is commonly referred to as ``spaghettification."
It is proposed in this paper that the tools
required to reduce the tidal forces to manageable 
levels can essentially eliminate the other 
obstacles, suggesting that traversable wormholes 
are indeed theoretically possible.  The seemingly 
highly restrictive low-tidal-force assumption 
can be justified by the charged-wormhole model 
due to Kim and Lee \cite{KL01}.  The same 
assumption is  used to study the stability of 
the wormhole by employing an equilibrium 
condition obtained from the 
Tolman-Oppenheimer-Volkoff equation.

Finally, the zero-tidal-force assumption 
enables us to invoke $f(R)$ modified gravity 
to eliminate the need for exotic matter near 
the throat.  The reason is that $f(R)$ modified 
gravity can be kept arbitrarily close  to 
Einstein gravity.

This paper is organized as follows:  Sec. 
\ref{S:structure} reviews the basic structure 
of a Morris-Thorne wormhole, while Sec. 
\ref{S:redshift} discusses the properties of 
the redshift and shape functions.  The 
zero-tidal-force assumption in Sec. \ref{S:zero} 
leads to the question of stability in Sec. 
\ref{S:stability}, followed by a discussion of 
$f(R)$ modified gravity and its consequences 
in Sec. \ref{S:f(R)}.  Sec. \ref{S:compatibility}    
considers the compatibility of the zero-tidal-force 
assumption with quantum field theory.  Sec. 
\ref{S:throat} deals with the problem of throat 
size.  Sec. \ref{S:low} then returns to the 
question of the feasibility of the low- or 
zero-tidal- force assumption, while Sec. 
\ref{S:emerge} discusses wormholes as emergent 
phenomena. 

\section{Wormhole structure}\label{S:structure}
Wormholes are handles or tunnels that could 
connect even widely separated regions of our 
Universe or different universes altogether.  
Wormholes seem to be as good a prediction of 
Einstein's theory as black holes, but, unlike 
the latter, they are subject to severe 
restrictions from quantum field theory.  For 
example, holding a wormhole open requires a 
violation of the null energy condition, which,
in turn, calls for the existence of ``exotic 
matter" in classical general relativity \cite 
{MT88}, a requirement that many researchers       
consider to be completely unphysical.

The line element for a Morris-Thorne wormhole 
is given by 
\begin{equation}\label{E:line1a}
  ds^{2}=-e^{2\Phi(r)}dt^{2}+\frac{dr^2}
  {1-\frac{b(r)}{r}}
  +r^{2}(d\theta^{2}+\text{sin}^{2}\theta\,
  d\phi^{2}),
  \end{equation}
using units in which $c=G=1$ .  
The motivation for this line element comes 
from Ref. \cite{MTW}:
\begin{equation*}
  ds^{2}=-e^{2\Phi(r)}dt^{2}+\frac{dr^2}
  {1-\frac{2m(r)}{r}}
  +r^{2}(d\theta^{2}+\text{sin}^{2}\theta\,
  d\phi^{2}),\quad r\le R
  \end{equation*}
  \begin{equation}\label{E:line1}
  =-\left(1-\frac{2M}{r}\right)dt^2
  +\frac{dr^2}{1-\frac{2M}{r}}
  +r^{2}(d\theta^{2}+\text{sin}^{2}\theta\,
  d\phi^{2}), \quad r>R.
\end{equation}
Here $m(r)$ is the effective mass inside
radius $r$, while $M$ is the mass of a star of
radius $R$ as seen by a distant observer.
If $\rho(r)$ is the energy density, then
the total mass-energy inside radius $r$
is given by
\begin{equation}
   m(r)=\int^r_04\pi (r')^2\rho(r')\,dr',
   \quad m(0)=0.
\end{equation}
In line element (\ref{E:line1a}), 
$\Phi=\Phi(r)$ is called the \emph{redshift
function}, which must be everywhere finite
to prevent the occurrence of an event horizon.  
The function $b=b(r)$ is called the 
\emph{shape function} since it determines the 
spatial shape of the wormhole when viewed, 
for example, in an embedding diagram \cite{MT88}.  
The spherical surface $r=r_0$ is called the 
\emph{throat} of the wormhole, where $b(r_0)=r_0$. 
 Additional requirements are 
$b'(r_0)<1$, called the \emph{flare-out
condition}, $b(r)<r$ for $r>r_0$,
and $b'(r_0)>0$.   Another
requirement is asymptotic flatness:
$\text{lim}_{r\rightarrow
\infty}\Phi(r)=0$ and $\text{lim}_{r\rightarrow
\infty}b(r)/r=0$.

It is noted in Ref. \cite{MT88} that 
the flare-out condition can only be met by
violating the null energy condition (NEC),
which states that
\begin{equation}
  T_{\alpha\beta}k^{\alpha}k^{\beta}\ge 0
\end{equation}
for all null vectors $k^{\alpha}$, where
$T_{\alpha\beta}$ is the energy-momentum
tensor.  As mentioned above, matter that 
violates the NEC is called ``exotic."  In
particular, for the outgoing null vector
$(1,1,0,0)$, the violation has the form
\begin{equation}\label{E:violation}
   T_{\alpha\beta}k^{\alpha}k^{\beta}=
   \rho +p_r<0.
\end{equation}
Here $T^t_{\phantom{tt}t}=-\rho$ is the
energy density, $T^r_{\phantom{rr}r}= p_r$
is the radial pressure, and
$T^\theta_{\phantom{\theta\theta}\theta}=
T^\phi_{\phantom{\phi\phi}\phi}=p_t$ is
the lateral (transverse) pressure.  Our
final task in this section is to list the 
Einstein field equations:

\begin{equation}\label{E:Einstein1}
  \rho(r)=\frac{b'}{8\pi r^2},
\end{equation}
\begin{equation}\label{E:Einstein2}
   p_r(r)=\frac{1}{8\pi}\left[-\frac{b}{r^3}+
   2\left(1-\frac{b}{r}\right)\frac{\Phi'}{r}
   \right],
\end{equation}
and
\begin{equation}\label{E:Einstein3}
   p_t(r)=\frac{1}{8\pi}\left(1-\frac{b}{r}\right)
   \left[\Phi''-\frac{b'r-b}{2r(r-b)}\Phi'
   +(\Phi')^2+\frac{\Phi'}{r}-
   \frac{b'r-b}{2r^2(r-b)}\right].
\end{equation}

\section{The redshift and shape functions}
    \label{S:redshift}
When discussing Morris-Thorne wormholes, it 
is generally assumed that $\Phi=\Phi(r)$ and 
$b=b(r)$ can be freely assigned and still 
retain a wormhole structure.  This assumption 
does not, however, address the practical 
problems of traversability, such as the tidal 
gravitational forces mentioned in the 
Introduction.  It is shown in Ref. \cite{MT88} 
that the radial tidal constraint is given by 
\begin{equation}
   \left|\left(1-\frac{b}{r}\right)
   \left(-\Phi''+\frac{b'r-b}{2r(r-b)}\Phi'
   -(\Phi')^2\right)\right|\le A,
\end{equation}    
where $A$ is a constant \cite{MT88}.  This 
constraint is most easily met if $\Phi'(r)
\equiv 0$ for all $r$, called the 
\emph{zero-tidal-force solution} in Ref. 
\cite{MT88}.  While a small value of 
$|\Phi'(r)|$ would suffice, we will retain 
the assumption $\Phi'(r)\equiv 0$ to keep 
the analysis tractable.  The implication 
of this assumption constitutes a major goal 
in this paper.  

It was noted in the Introduction that we can 
hypothesize a basic wormhole structure -- and 
this would include the existence of a shape 
function.  This shape function must meet 
certain requirements, as discussed in the 
previous section.  To address this issue, 
it is proposed in Ref. \cite{pK20a} that 
a generic shape function can be defined 
by starting with the following family:
\begin{equation}\label{E:shape}
   b_{\eta}(r)=r_0\left(\frac{r}{r_0}
      \right)^{1-\eta},\quad 0<\eta<1.
\end{equation}
Evidently, $b_{\eta}(r_0)=r_0$, while
\begin{equation}\label{E:bprime}
    0<b'_{\eta}(r_0)=1-\eta<1.
\end{equation}
For this program to work, we need to 
assume that a typical shape function 
$b=b(r)$ is concave down in the immediate 
vicinity of $r=r_0$ with $b(r)<r$.  More 
precisely, we require that
(a) $b(r_0)=r_0,$
(b) $0<b'(r_0)<1,$ and
(c) $b(r)$ is concave down near $r=r_0$.
Properties (a) and (b) are clearly met.  
Regarding Property (c), for any
$\eta\,\,\varepsilon\, (0,1)$,
\[
    b''_{\eta}(r_0)=\frac{1}{r_0}(1-\eta)(-\eta)
    <0,
\]
showing that $b_{\eta}(r)$ is indeed concave
down near $r=r_0$.  We conclude that any 
shape function that meets Properties (a), 
(b), and (c) can be approximated by some 
$b_{\eta}(r)$ in the vicinity of the throat.
For example, the special case $\eta=1/2$ 
leads to the parabola $b(r)=\sqrt{r_0r}$, 
which is concave down to the right of 
$r=r_0$ with $b(r)<r$ for all $r$. The 
same behavior is exhibited by all members 
of the family in Eq. (\ref{E:shape}). 

In summary, for every value $a$ between 
0 and 1, there exists a member 
$b_{\eta}(r)$ such that $b'_{\eta}(r_0)=a$.

\section{The zero-tidal-force assumption}     
     \label{S:zero}
The assumption $\Phi'(r)\equiv 0$ was 
introduced as a necessary traversability 
condition.  The rest of this paper will 
be devoted to showing that the purpose 
of this assumption goes far beyond these 
considerations.  In particular, Sec. 
\ref{S:stability}, \ref{S:f(R)}, and 
\ref{S:compatibility} discuss, respectively, 
the question of stability, the feasibility 
of invoking $f(R)$ modified gravity to 
eliminate the need for exotic matter, 
and the compatibility of traversable 
wormholes with quantum field theory.  We 
are also going to conclude that 
Morris-Thorne wormholes can only exist 
for relatively large throat sizes.

\section{Stability}\label{S:stability}    
Stability is an important topic in 
wormhole physics.  The special case of 
zero-tidal-force wormholes is taken up 
in Ref. \cite{pK20b} by employing an 
equilibrium condition obtained from the 
Tolman-Oppenheimer-Volkoff (TOV) equation 
\cite{jP93, RKRI}.  
\begin{equation}\label{E:TOV}
   \frac{dp_r}{dr}+\Phi'(\rho+p_r)+
   \frac{2}{r}(p_r-p_t)=0.
\end{equation}
The equilibrium state of a structure is
determined from the three terms in this
equation, defined as follows: the
gravitational force
\begin{equation}
   F_g=-\Phi'(\rho+p_r),
\end{equation}
the hydrostatic force
\begin{equation}\label{E:hydrostatic}
   F_h=-\frac{dp_r}{dr},
\end{equation}
and the anisotropic force
\begin{equation}\label{E:anisotropic}
   F_a=\frac{2(p_t-p_r)}{r}
\end{equation}
due to the anisotropic pressure in a
Morris-Thorne wormhole.
Eq. (\ref{E:TOV}) then yields the
following equilibrium condition:
$F_g+F_h+F_a=0.$  Since $\Phi'\equiv 0$, 
the equilibrium condition becomes
\begin{equation}
   F_h+F_a=0.
\end{equation}
To show that this condition is met in 
this paper, we first need to consider 
$f(R)$ modified gravity.

\section{$f(R)$ modified gravity}
     \label{S:f(R)}
The purpose of this section is two-fold, 
to introduce $f(R)$ modified gravity and 
to show that we can remain arbitrarily 
close to Einstein gravity.

Here we need to retain our assumption 
$\Phi'(r)\equiv 0$.  Otherwise, according 
to Lobo and Oliveira \cite{LO09}, the 
analysis becomes intractable.  Fortunately, 
this requirement is in line with our 
overall goals, as we saw in Sec. 
\ref{S:zero}. 

Next, we list the gravitational field 
equations in the form given in Ref.
\cite{LO09}:
\begin{equation}\label{E:Lobo1}
   \rho(r)=F(r)\frac{b'(r)}{r^2},
\end{equation}
\begin{equation}\label{E:Lobo2}
   p_r(r)=-F(r)\frac{b(r)}{r^3}
   +F'(r)\frac{rb'(r)-b(r)}{2r^2}
   -F''(r)\left(1-\frac{b(r)}{r}\right),
\end{equation}
and
\begin{equation}\label{E:Lobo3}
   p_t(r)=-\frac{F'(r)}{r}\left(1-\frac{b(r)}{r}
   \right)+\frac{F(r)}{2r^3}[b(r)-rb'(r)],
\end{equation}
where $F=\frac{df}{dR}$.  The Ricci
curvature scalar is given by
\begin{equation}\label{E:R}
   R(r)=\frac{2b'(r)}{r^2}.
\end{equation}

To see the connection to the flare-out
condition at the throat, observe that from
Eqs. (\ref{E:violation}), (\ref{E:Einstein1}),
and (\ref{E:Einstein2}), we have
\begin{equation*}\label{E:exotic}
   8\pi[\rho(r_0)+p_r(r_0)]=\frac{r_0b'(r_0)
   -b(r_0)}{r_0^3}<0   
\end{equation*}
since $b(r_0)=r_0$.  Given that the radial 
tension $\tau(r)$ is the negative of 
$p_r(r)$, Eq. (\ref{E:violation}) implies
that $\tau-\rho c^2>0$, temporarily 
reintroducing $c$.  The last inequality 
has given rise to the designation
``exotic matter" since $\tau>\rho c^2$
implies that there is an enormous radial
tension at the throat.  

In this paper, we are going to choose
\begin{equation}\label{E:modified}
   f(R)=aR^{1\pm\epsilon},\quad
       \epsilon\ll 1,
\end{equation}
where $a$ is a constant.  The reason is 
that since $\epsilon$ can be arbitrarily 
close to zero, the resulting $f(R)$ 
modified gravity can be arbitrarily 
close to Einstein gravity.  Since 
$F=\frac{df}{dR}$, we get from Eq.
(\ref{E:R})
\begin{equation}\label{E:F}
   F=a(1\pm\epsilon)R^{\pm\epsilon}=
   a(1\pm\epsilon)\left(\frac{2b'(r)}
   {r^2}\right)^{\pm\epsilon}.
\end{equation}
This is enough to show that $F_h+F_a=0$:
\begin{multline}
   F_h+F_a=\\a(1\pm\epsilon)
   \left[2^{\pm\epsilon}\left(\frac{\rho}
   {2^{\pm\epsilon}a(1\pm\epsilon)}
   \right)^{\frac{\pm\epsilon}{1\pm\epsilon}}\right]
    \left(\frac{rb'(r)-3b(r)}{r^4}+
    \frac{b(r)-rb'(r)}{r^4}+\frac{2b(r)}{r^4}
    \right)=0.
\end{multline}
(See Ref. \cite{pK20b} for details.)  We 
now see that the equilibrium condition 
is satisfied, thereby yielding a stable 
wormhole.  This outcome leads to one of 
our most important conclusions: since
our modified theory, based on Eq. 
(\ref{E:modified}), can be arbitrarily 
close to Einstein's theory, the stability 
criterion carries over to Morris-Thorne 
wormholes.  

Returning to Ref. \cite{LO09}, it is 
shown that for the material threading 
the wormhole, the NEC can be met, 
thereby allowing the use of ordinary 
(nonexotic) matter in the modified theory, 
even if it is arbitrarily close to 
Einstein's theory.  As a first step, by 
imposing the conditions $\rho+p_r\ge 0$ 
and $\rho\ge 0$, it now follows from 
Eqs. (\ref{E:Lobo1}) and (\ref{E:Lobo2})  
that the function $F$ must be positive 
and that it must satisfy the following 
conditions at the throat:
\begin{equation}
   \frac{Fb'}{r^2}\ge 0
\end{equation}
and 
\begin{equation}
   \frac{(2F+rF')(b'r-b)}{2r^3}-
      F''\left(1-\frac{b}{r}\right)\ge 0.
\end{equation}
From Eq. (\ref{E:Lobo3}), we also have 
$\rho+p_t\ge 0$ at the throat, $F$ being 
positive.  To complete the proof of the 
above assertion, it is shown in Ref. 
\cite{pK21} that the NEC is met for all 
null vectors $(1,a,b,c)$, where 
$0 \le a, b,c \le 1$ and $a^2+b^2+c^2=1.$

\section{Compatibility with quantum field theory}
    \label{S:compatibility}
We have seen in this paper that the 
zero-tidal-force assumption plays a critical 
role.  This section continues the theme by 
examining how this assumption affects the 
compatibility of classical wormhole theory 
with quantum field theory, which places some 
severe constraints on Morris-Thorne wormholes 
\cite{FR96}.  More precisely, the wormhole 
spacetime must satisfy a certain quantum 
inequality in an inertial Minkowski spacetime 
without boundary: if $u^{\mu}$ is the observer's 
four-velocity and  $\langle T_{\mu\nu}u^{\mu}u^{\nu}\rangle$
is the expected value of the local energy
density in the observer's frame of reference,
then
\begin{equation}\label{E:QFT}
   \frac{\tau_0}{\pi}\int^{\infty}_{-\infty}
   \frac{\langle T_{\mu\nu}u^{\mu}u^{\nu}\rangle
   d\tau}{\tau^2+\tau_0^2}\ge
   -\frac{3}{32\pi^2\tau_0^4},
\end{equation}
where $\tau$ is the observer's proper time
and $\tau_0$ the duration of the sampling time.
(See Ref. \cite{FR96} for details.)  The
inequality can be applied in a curved
spacetime as long as $\tau_0$ is small
compared to the local proper radius of
curvature.  The desired estimates of the
local curvature are obtained from the
components of the Riemann curvature tensor
in classical general relativity.  It 
follows that the exotic matter must be 
confined to a narrow band around the 
throat \cite{FR96}.  Now, according to 
Refs. \cite{pK08, pK09}, this can only 
be accomplished by fine-tuning the 
metric coefficients.  In other words, 
to satisfy the quantum inequality, one 
must strike a balance between reducing 
the size of the exotic region and the 
degree of fine-tuning of the metric 
coefficients required to achieve this 
reduction.  As a result, $\Phi'(r)$ 
must be fine-tuned to remain in a 
narrow range.  The most important 
conclusion for our purposes is that 
$\Phi'(r)\equiv 0$ is outside this 
range, so that the resulting wormhole 
solution cannot be compatible with 
quantum field theory.  This also 
applies to the wormhole solutions 
in Ref. \cite{MT88}.    

Now the reason for invoking $f(R)$ 
modified gravity becomes apparent: the
estimates of the local curvature needed to
apply Inequality (\ref{E:QFT}) come from
Einstein's theory, not from the modified
theory.  So the previous objections do
not apply.  More precisely, in the
equation $f(R)=aR^{1\pm\epsilon}$,
$\epsilon$ is always nonzero.  So
even if the modified theory is
arbitrarily close to Einstein's theory,
it remains an $f(R)$ theory, thereby
avoiding a direct conflict with
quantum field theory.

\section{The throat size $r=r_0$}
     \label{S:throat}
     
We saw in Sec. \ref{S:f(R)} that $f(R)$ 
modified gravity can remain arbitrarily 
close to Einstein gravity.  In this 
section a slight shift in emphasis 
yields a slightly modified theory that 
enables us to estimate the throat size 
$r=r_0$ of a Morris-Thorne wormhole.  
To that end, observe that the field 
equations (\ref{E:Lobo1})-(\ref{E:Lobo3})     
reduce to the Einstein equations for 
$\Phi'(r)\equiv 0$ whenever $F(r)\equiv 1$.
Next, from Eqs. (\ref{E:Lobo1}) and 
(\ref{E:R}), we have 
\begin{equation}
   F(r)=\frac{2\rho(r)}{2\frac{b'(r)}{r^2}}=
   \frac{2\rho(r)}{R(r)}.
\end{equation}
So a slight change in $F$ results in a 
slight change in $R$, enough to quantify 
the notion of slightly modified gravity: 
assume that $F(r)$ remains close to unity 
and relatively ``flat," i.e., both $F'(r)$ 
and $F''(r)$ remain relatively small in 
absolute value.

We already saw in Sec. \ref{S:f(R)} the 
$\rho+p_r\ge 0$ in $f(R)$ modified gravity. 
So from Eqs. (\ref{E:Lobo1}) and 
(\ref{E:Lobo2}) 
\begin{equation}\label{E:positive}
   \rho+p_r=(rb'-b)\left(\frac{F}{r^3}
   +\frac{F'}{2r^2}\right)-F''\left(1-\frac{b}{r}
   \right)\ge 0.
\end{equation}
To draw our conclusion, it is convenient to 
use a simple example: suppose 
$F=2-e^{a(r-r_0)}$; then $F'=-e^{a(r-r_0)}a$,
and $F''=-e^{a(r-r_0)}a^2$, where $a$ is a 
small constant.  At $r=r_0$, $F(r_0)=1$, 
$F'(r_0)=-a$, and $F''(r_0)=-a^2$.  Substituting 
in Eq. (\ref{E:positive}), we get
\begin{equation}
   (rb'-b)\frac{2-ar_0}{2r_0^3}+
   a^2\left(1-\frac{b}{r}\right)\ge 0
\end{equation}
at $r=r_0$, provided that 
\begin{equation}
   r_0\ge \frac{2}{a}.
\end{equation}
Given that $a$ can be extremely small, we 
get a valid wormhole solution only for 
sufficiently large throat sizes.  (For 
further details, see Ref. \cite{pK13a}.)

This result is consistent with a problem 
already discussed in Ref. \cite{MT88}, 
the radial tension at the throat.  First 
we need to recall that the radial tension 
$\tau(r)$ is the negative of the radial 
pressure $p_r(r)$.  According to Ref. 
\cite{MT88}, the Einstein field equations 
can be rearranged to yield $\tau(r)$.  
Temporarily reintroducing $c$ and $G$, 
we obtain
\begin{equation}
   \tau(r)=\frac{b(r)/r-2[r-b(r)]\Phi'(r)}
   {8\pi Gc^{-4}r^2}.
\end{equation}
The radial tension at the throat therefore 
becomes
\begin{equation}\label{E:tau}
  \tau(r_0)=\frac{1}{8\pi Gc^{-4}r_0^2}\approx
   5\times 10^{41}\frac{\text{dyn}}{\text{cm}^2}
   \left(\frac{10\,\text{m}}{r_0}\right)^2.
\end{equation}
As noted in Ref. \cite{MT88}, for a throat size 
of $r_0=3$ km, $\tau(r)$ has the same magnitude
as the pressure at the center of a massive
neutron star.  Eq. (\ref{E:tau}) shows that 
Morris-Thorne wormholes could only exist on 
very large scales.

\emph{Remark:} The above discussion suggests 
that moderately-sized wormholes are actually 
compact stellar objects \cite{pK22} and are 
thereby beyond the scope of the present study.

\section{Feasibility: wormholes with low 
   tidal forces}\label{S:low}
   
Our conclusions so far depended on the 
zero-tidal-force assumption $\Phi'(r)
\equiv 0$.  It was noted in Sec. 
\ref{S:redshift}, however, that a small 
value of $|\Phi'(r)|$ would have been 
sufficient, thereby remaining the single 
most important condition.  In this section 
we will examine the circumstances under 
which this condition can be satisfied.  
This will naturally call for some 
requirements beyond those already considered.

We start by assuming a noncommutative-geometry 
background: an important outcome of string 
theory is the realization that coordinates 
may become noncommutative operators on a 
$D$-brane \cite{eW96, SW99}.  The result is 
a fundamental discretization of spacetime 
due to the commutator 
$[\textbf{x}^{\mu},\textbf{x}^{\nu}]
=i\theta^{\mu\nu}$, where $\theta^{\mu\nu}$
is an antisymmetric matrix.  According to 
Refs. \cite{SS03, NSS06, mR11}, noncommutativity 
replaces point-like objects by smeared 
objects, a procedure that is consistent with 
the Heisenberg uncertainty principle.  The 
purpose is to eliminate the divergences 
that normally occur in general relativity.

A common way to model the smearing is by 
means of a Gaussian distribution of minimal 
length $\sqrt{\alpha}$ instead of the obvious 
alternative, the Dirac delta function 
\cite{fR12, pK13b, LL12, NM08}.  A simpler 
but equally effective way is to assume 
that the energy density of a static and 
spherically symmetric and particle-like 
gravitational source has the form 
\begin{equation}\label{E:rho}
  \rho(r)=\frac{m\sqrt{\alpha}}
     {\pi^2(r^2+\alpha)^2}.
\end{equation}
(See Refs. \cite{fR12} and \cite{ pK13b}.)  
The basic idea is that the mass $m$ is 
diffused throughout the region of linear 
dimension $\sqrt{\alpha}$ due to the 
uncertainty.  It is emphasized in Ref. 
\cite{NSS06} that noncommutative geometry 
is an intrinsic property of spacetime 
and does not depend on any particular 
feature such as curvature.  Furthermore, 
to make use of Eq. (\ref{E:rho}), we can 
keep the standard form of the Einstein 
field equations in the sense that the 
Einstein tensor retains its original form, 
but the stress-energy tensor is modified 
\cite{NSS06}.  It follows that the length 
scales can be macroscopic.

Next, given that black holes can carry an 
electric charge, it is natural to assume 
that wormholes can do likewise.  It is 
proposed by Kim and Lee \cite{KL01} that 
for a wormhole with constant electric 
charge $Q$, the Einstein field equations 
are given by 
\begin{equation}\label{E:KL01}
   G^{(0)}_{\mu\nu}+G^{(1)}_{\mu\nu}=
   8\pi[T^{(0)}_{\mu\nu}+T^{(1)}_{\mu\nu}].
\end{equation}  
According to Ref. \cite{KL01}, since the 
usual form is $G^{(0)}_{\mu\nu}=8\pi
T^{(0)}_{\mu\nu}$, the modified form, 
Eq. (\ref{E:KL01}), is obtained by adding 
the matter term $T^{(1)}_{\mu\nu}$ to the 
right side and the corresponding back 
reaction term $G^{(1)}_{\mu\nu}$ to the 
left side.  The proposed metric then becomes 
\begin{equation}\label{E:line3}
  ds^{2}=-\left(1+\frac{Q^2}{r^2}\right)dt^{2}
  +\left(1-\frac{b(r)}{r}+\frac{Q^2}{r^2}
  \right)^{-1}dr^2
  +r^{2}(d\theta^{2}+\text{sin}^{2}\theta\,
  d\phi^{2}).
\end{equation}
It is shown in Ref. \cite{KL01} that this 
metric is a self-consistent solution to the 
Einstein field equations.  It also follows
that the (effective) shape function is 
given by 
\begin{equation}
   b_{\text{eff}}(r)=b(r)-\frac{Q^2}{r}.
\end{equation}
In the Kim-Lee model, the total charge is 
given by $\int\int\int_V \rho_q (r) dV$, 
where $\rho_q$ is the charge density.  To 
adapt this to our noncommutative-geometry 
background, it is proposed in Ref. 
\cite{KG17} that
\begin{equation}
   \rho_q(r)=\frac{Q^2\sqrt{\alpha}}
   {\pi^2(r^2+\alpha)^2},
\end{equation}
where $Q^2$ refers to the Kim-Lee model.  
So the smeared charge $Q_{\alpha}^2(r)$ is 
\begin{equation}
   Q^2_{\alpha}(r)=\int^r_04\pi(r')^2
   \frac{Q^2\sqrt{\alpha}}{\pi^2
   [(r')^2+\alpha]^2}dr'=
   \frac{2Q^2\sqrt{\alpha}}{\pi}
   \left(\frac{1}{\sqrt{\alpha}}\text{tan}^{-1}
   \frac{r}{\sqrt{\alpha}}-
   \frac{r}{r^2+\alpha}\right).
\end{equation}
It is shown in Ref. \cite{KG17} that the 
shape function is 
\begin{multline}
   b_{\text{eff}(r)}=\frac{4m\sqrt{\alpha}}{\pi}
    \left(\frac{1}{\sqrt{\alpha}}\text{tan}^{-1}
   \frac{r}{\sqrt{\alpha}}-
   \frac{r}{r^2+\alpha}\right)
   -\frac{1}{r}\frac{2Q^2\sqrt{\alpha}}{\pi}
    \left(\frac{1}{\sqrt{\alpha}}\text{tan}^{-1}
   \frac{r}{\sqrt{\alpha}}-
   \frac{r}{r^2+\alpha}\right)\\
   -\frac{4m\sqrt{\alpha}}{\pi}
    \left(\frac{1}{\sqrt{\alpha}}\text{tan}^{-1}
   \frac{r_0}{\sqrt{\alpha}}-
   \frac{r_0}{r_0^2+\alpha}\right)+r_0.
\end{multline}
Moreover, $b_{\text{eff}(r)}$ satisfies all 
the requirements of a shape function.

We can now turn to our main goal, estimating 
$|\Phi'(r)|$.  From line element ({\ref
{E:line3}), 
\begin{equation}
   e^{2\Phi}=1+\frac{Q^2}{r^2},
\end{equation}
whence
\begin{equation}
   \Phi(r)=\frac{1}{2}\text{ln}\left[1+
   \frac{1}{r^2}\frac{2Q^2\sqrt{\alpha}}{\pi}
   \left(\frac{1}{\sqrt{\alpha}}\text{tan}^{-1}
   \frac{r}{\sqrt{\alpha}}-\frac{r}{r^2+\alpha}
   \right)\right]
\end{equation}
and 
\begin{equation}
   \Phi'(r)=\frac{-\frac{1}{r^3}\frac{2Q^2\sqrt{\alpha}}
   {\pi}\left(\frac{1}{\sqrt{\alpha}}\text{tan}^{-1}
   \frac{r}{\sqrt{\alpha}}-\frac{r}{r^2+\alpha}\right)
   +\frac{2Q^2\sqrt{\alpha}}{\pi(r^2+\alpha)^2}}
   {1+\frac{1}{r^2}\frac{2Q^2\sqrt{\alpha}}{\pi}
   \left(\frac{1}{\sqrt{\alpha}}\text{tan}^{-1}
   \frac{r}{\sqrt{\alpha}}-\frac{r}{r^2+\alpha}\right)}.
\end{equation}
Since $\alpha$ is a small parameter, the easiest 
way to estimate $|\Phi'(r)|$ is to let 
$\alpha\rightarrow 0$:
\begin{equation}
   |\Phi'(r)|\approx\left|\frac{-\frac{1}{r^3}Q^2}
   {1+\frac{1}{r^2}Q^2}\right|.
\end{equation}
We saw in Sec. \ref{S:throat} that we get a valid 
wormhole solution only for sufficiently large 
throat sizes, showing that $|\Phi'(r)|$ is indeed 
arbitrarily small, thereby overcoming the restrictions 
from quantum field theory. 

\section{Feasibility: wormholes as emergent 
    phenomena}\label{S:emerge}
   
It was noted in the Introduction that we 
can assume that a basic wormhole structure 
may be hypothesized.  We will now strengthen 
this assumption by showing that wormholes can 
be viewed as emergent phenomena.  Some aspects 
of this problem have already been discussed in 
Ref. \cite{pK23a}. 

The basic idea of emergence is that certain 
high-level phenomena cannot be deduced even 
in principle from a lower-level domain.  For 
example, ant colonies are capable of building 
extremely complex structures, but this outcome 
cannot be explained by examining the behavior 
of individual ants.  This behavior is an 
example of a \emph{fundamental property} and 
the structure an example of an \emph{emergent 
property}.  As another example, life emerges 
from objects that are themselves completely 
lifeless, such as atoms and molecules.  Such 
a process is not reversible, however, even in 
principle: living organisms tell us nothing 
about the particles in the fundamental theory.  
In fact, these properties are not even relevant 
in the resulting \emph{effective model}, thereby 
illustrating the often surprising or unexpected 
outcomes that characterize emergent phenomena.  
Moreover, noncommutative geometry in the form 
discussed in the previous section is an example 
of a fundamental property, thereby calling 
attention to the fact that quantum mechanics 
generally incorporates many such fundamental 
phenomena.  More precisely, it is argued 
in Ref. \cite{pL17} that in this context, 
emergence is based on entanglement.  For 
our purposes we simply acknowledge that 
two entangled particles are connected by 
a type of wormhole called an Einstein-Rosen 
bridge \cite{MS13}.  This can now be seen as 
a fundamental property, while the emergent 
property is a macroscopic wormhole.  Of 
course, such wormholes have never been 
observed, but the very possibility of 
emergence seems to greatly increase the 
probability that such wormholes actually 
exist. 

This observation is made more concrete in 
Ref. \cite{pK23b}.  Using Eq. (\ref{E:rho}), 
the shape function becomes         
\begin{multline}
   b(r)=r_0+\int^r_{r_0}8\pi(r')^2\rho(r')dr'\\
   =\frac{4m}{\pi}
  \left[\text{tan}^{-1}\frac{r}{\sqrt{\beta}}
  -\sqrt{\beta}\frac{r}{r^2+\beta}-
  \text{tan}^{-1}\frac{r_0}{\sqrt{\beta}}
  +\sqrt{\beta}\frac{r_0}{r_0^2
  +\beta}\right]+r_0\\
  =\frac{4m}{\pi}\frac{1}{r}
  \left[r\,\text{tan}^{-1}\frac{r}{\sqrt{\beta}}
  -\sqrt{\beta}\frac{r^2}{r^2+\beta}-
  r\,\text{tan}^{-1}\frac{r_0}{\sqrt{\beta}}
  +\sqrt{\beta}\frac{r_0r}{r_0^2
  +\beta}\right]+r_0.
\end{multline}
The reason is that we can now define a new 
shape function $B=b/\sqrt{\beta}$:
\begin{multline}
   \frac{1}{\sqrt{\beta}}
   \,b(r)=
   B\left(\frac{r}{\sqrt{\beta}}\right)=\\
   \frac{4m}{\pi}\frac{1}{r}\left[\frac{r}{\sqrt{\beta}}
   \,\text{tan}^{-1}\frac{r}{\sqrt{\beta}}
   -\frac{\left(\frac{r}{\sqrt{\beta}}\right)^2}
   {\left(\frac{r}{\sqrt{\beta}}\right)^2+1}
  -\frac{r}{\sqrt{\beta}}\,
  \text{tan}^{-1}\frac{r_0}{\sqrt{\beta}}
  +\frac{r}{\sqrt{\beta}}
  \frac{\frac{r_0}{\sqrt{\beta}}}
  {\left(\frac{r_0}{\sqrt{\beta}}\right)^2+1}
  \right]+\frac{r_0}{\sqrt{\beta}}.
\end{multline}
It follows that
\begin{equation}
   B\left(\frac{r_0}{\sqrt{\beta}}\right)
   =\frac{r_0}{\sqrt{\beta}},
\end{equation}
the analogue of $b(r_0)=r_0$.  So the 
throat size can be macroscopic, confirming 
the claim that we are indeed dealing with 
an emergent property.

\section{Conclusion}
Geometrically speaking, wormholes may  
be compared to black holes, but they 
are subject to severe restrictions from 
quantum field theory.  This raises some 
questions regarding the number and severity 
of these constraints, as well as conditions 
needed to allow a wormhole to be used for 
interstellar travel.  It is proposed in this 
paper that controlling the radial tidal forces, 
usually associated with black holes, can 
essentially eliminate the other obstacles.  
To keep the analysis tractable, we assume 
zero tidal forces (Sec. \ref{S:zero}), an 
assumption that immediately yields a stable 
solution (Sec. \ref{S:stability}).  The same 
assumption enables us to invoke $f(R)$ 
modified gravity (Sec. \ref{S:f(R)}), thereby 
eliminating the need for exotic matter at the 
thoat of a Morris-Thorne wormhole.  The reason 
is that $f(R)$ modified gravity can be kept 
arbitrarily close to Einstein gravity 
(Sec. \ref{S:f(R)}).

It is shown in Sec. \ref{S:compatibility} 
that the zero-tidal-force assumption would 
make the wormhole solution incompatible with 
quantum field theory in classical general 
relativity but would be acceptable in $f(R)$ 
modified gravity. 

It is  noted in Sec. \ref{S:introduction} that 
we may hypothesize a basic wormhole structure  
(discussed in Sec. \ref{S:structure}), 
which necessarily includes a shape function.  
Moreover, it is pointed out in Sec. 
\ref{S:redshift} that any valid shape 
function can be approximated by a member 
of the family $b_{\eta}(r)
=r_0(r/r_0)^{1-\eta},\, 0<\eta<1$,  
and in Sec. \ref{S:throat} that a 
Morris-Thorne wormhole can only exist on 
very large scales, i.e., with a large 
$r=r_0$, which is consistent with the 
discussion following Eq. (\ref{E:tau}).  

Finally, Sec. \ref{S:low} discusses the 
feasibility of this approach by introducing 
several additional conditions to ensure 
that the radial tidal forces can indeed be 
kept sufficiently low, while Sec. \ref{S:emerge} 
discusses wormholes as emergent phenomena.

\end{document}